\documentclass[universe,article,accept,pdftex,moreauthors]{Definitions/mdpi} 

\firstpage{1} 
\pubvolume{1}
\issuenum{1}
\articlenumber{0}
\pubyear{2025}
\copyrightyear{2025}
\externaleditor{~} 
\datereceived{30 May 2025} 
\daterevised{23 July 2025} 
\dateaccepted{11 August 2025} 
\datepublished{ } 

\hreflink{https://doi.org/} 

\Title{Powerful Radio Sources as Probes of Black Hole Physics}

\TitleCitation{Powerful Radio Sources as Probes of Black Hole Physics}

\Author{{Ruth} 
{ A.} Daly \orcidA{0000-0002-5503-0691}}

\AuthorNames{Ruth A. Daly}

\isAPAStyle{%
       \AuthorCitation{Daly, Ruth A.}
         }{%
        \isChicagoStyle{%
        \AuthorCitation{Daly, Ruth A.}
        }{
        \AuthorCitation{Daly, R.A.}
        }
}

\address[1]{{Department} 
 of Physics, {Penn State University, Berks Campus} 
, Reading, PA {19610}
, USA; rdaly@psu.edu}

\abstract{Powerful jetted radio sources for which the luminosity in directed kinetic energy has been empirically determined, independent of assumptions, are considered. The total outflow lifetime of each source determined in the context of detailed cosmological studies was found to depend only upon the luminosity in directed kinetic energy ($L$). The distributions of $L$, total outflow lifetime, and total outflow energy each have a broad range of values, as do the supermassive black hole masses. The total outflow energy relative to the black hole mass is a small number with a small dispersion. Three explanations of these remarkable results are considered. This could indicate (1) the efficiencies with which black hole irreducible mass is increased and spin mass energy is extracted during the outflow event, (2) that the merger of two supermassive black holes occurs over a timescale commensurate with the independently determined outflow lifetime and that these mergers lead to the production of the low-frequency gravitational wave background, or (3) that feedback shuts off black hole accretion due to energy injected into the ambient medium.}

\begin{document}
\section{Introduction}

The ``mass'' of an astrophysical black hole typically refers to the total or dynamical black hole mass. This is the mass measured by a distant observer using standard astronomical techniques. Another important property of an astrophysical black hole is the black hole spin. The total dynamical mass ($M$) of a spinning black hole has contributions from the ``irreducible'' black hole mass ($M_{irr}$) and the ``rotational'' black hole mass ($M_{rot}$, 
where $M^2 = M_{irr}^2 + M_{rot}^2$,  with   $M_{rot} \equiv Jc/(2GM_{irr})$, where $J$ is the spin angular momentum of the black hole)~\cite{C70,CR71,H71}. A fourth important quantity is the spin mass energy of the black hole that is available to be extracted, i.e., $M_{spin} = M - M_{irr}$~\cite{P69, PF71}. 

For astrophysical black holes, the spin mass energy ($M_{spin}$) is a critically important parameter, since it indicates the mass energy that is available to be extracted from the black hole. When spin mass energy is extracted from a black hole, the black hole's irreducible mass must increase during the extraction, since the extraction is never 100\% \mbox{efficient~\cite{Ketal21,Setal21,Tetal21,RR23,RR24,R25}.} Different spin mass-energy extraction  processes that could power jetted outflows from black holes have different efficiency factors that describe the amount of spin mass energy extracted and the increase in the black hole's irreducible mass during the extraction process. Processes involving electric and magnetic fields are among the most efficient \mbox{processes~\cite{Ketal21,Setal21,Tetal21,RR23,RR24,R25}.} 

Jetted outflows from supermassive and stellar-mass black holes are ubiquitous in the universe. The outflows are observed to be magnetized, and the jetted outflows are thought to be produced as the result of processes that involve strong---that is, 
dynamically important---magnetic fields. The interaction between a jetted outflow and the ambient gas occurs through collisionless shocks, which exist due to the presence of magnetic fields~\cite{H85, Daly1990, BMR19}. Indeed, it seems reasonable to suppose that the recently detected ``Little Red Dots'' could represent the state of supermassive black holes before the introduction of magnetic 
fields into the ambient gaseous medium by stellar winds and other processes---that is, the fact that no radio emission  has been detected from Little Red Dots~\cite{Ketal25,Retal25,Petal25, Getal25,Letal25} could indicate that these young systems have not yet built up sufficiently strong magnetic fields to allow for the production of detectable radio emissions. 

The connections between black hole physical processes and properties and 
the properties of powerful jetted outflows from black hole systems are discussed here. The sources that are considered are very powerful classical double-radio sources, as described in Section~\ref{TheRadioSources}. The application of the source properties to understand the processes associated with jet production for these sources and the impact of jet production on the black hole mass components ($M_{spin}$, $M_{irr}$, and $M_{rot}$) are discussed in Section~\ref{Connections}. In this section, three ways that the properties of jetted outflows  can be applied to understand the black hole physics relevant to the production of jets are considered in detail, as are 
their implications. The results and conclusions are summarized and discussed in Section \ref{SummaryConclusions}. Unless otherwise specified, all results were obtained in the context of a standard Lambda Cold Dark Matter (LCDM) model with $\Omega_{\Lambda} =0.7$, $\Omega_m = 0.3$, and a Hubble constant of $100~ h$ {km} 
/s/Mpc with a value of $h = 0.7$. 

\section{The Radio Sources}
\label{TheRadioSources}
There are numerous types of extended radio sources that are produced by dual-jetted outflows associated with 
supermassive black hole systems, including FRI and 
FRII sources~\cite{FR74,DY76,M80}.  
The classification of large extended radio sources into two categories was proposed by Fanaroff and Riley in 1974~\cite{FR74}, which are, thus, referred to as ``FRI'' and ``FRII'' sources.
The radio emission is produced by a population of relativistic electrons in the presence of a magnetic field. 
FRI and FRII extended radio sources have a broad range of radio power and a broad range of morphological characteristics. The 
``supermassive black hole system'' is taken to include the black hole; the material in the immediate vicinity of the black hole, which is referred to as the ``accretion disk''; and the dual-jetted outflow. The dual-jetted outflow produces an extended radio source, that is, a radio source that is as large as or larger than the 
host galaxy.  

FRI sources are ``edge-darkened'', and FRII sources are ``edge-brightened''~\cite{FR74}. FRII sources are characterized by relatively small, bright radio ``hotspots'' located a significant distance from the host supermassive black hole. Typically, there are two jets emanating roughly symmetrically from the location of the host supermassive black hole, which are referred to as ``dual jets.'' 

FRII sources, also known as ``classical doubles,''  subdivide into numerous categories. {Ref.} 
 \cite{LW84} identified five primary 
and three secondary types of FRII radio-source structure, for a total of eight FRII source types. 
For example, {Refs.} \cite{LW84, AL87, LMS89} refer to the extended 
radio-emitting region that lies between the host supermassive black hole and each of the two ends of the extended radio source as the ``radio bridge'', and this nomenclature is adopted here. Note that the ``radio bridge'' is sometimes referred to as ``radio lobes.'' 

Of the five primary and three secondary types of FRII radio bridge structure 
\cite{LW84}, only one exhibits a long, straight radio bridge with no distortions. This the type 1 FRII source. These type 1 FRII sources exhibit 
a quite regular, cylindrical, ``cigar-shaped'' radio-bridge structure. Each of the remaining seven
FRII categories exhibits various types of radio-bridge distortions. Following up on this, {Ref.} \cite{D02}
defined FRIIa sources as those  
with a radio-bridge distortion of any type and FRIIb sources as those with regular, cylindrical, cigar-shaped radio bridges---that is, type 1 FRII sources are referred to here as FRIIb sources, and the seven remaining FRII types defined by~\cite{LW84} are referred to as FRIIa sources. 

All of the radio sources discussed in this paper are FRIIb radio sources with a regular ``cigar-shaped'' radio bridge structure. It was found that the FRII type is strongly  correlated  
with the radio power and that only the most powerful radio sources exhibited very regular ``cigar-shaped'' radio 
bridges, that is, are FRIIb sources  
\cite{LMS89, Daly2010}. The characteristics of FRIIb sources 
make them ideal for the studies presented here, and their properties are described in Section 
\ref{FRIIb}. The radio-source properties indicate that the forward region of the source is moving into 
the ambient gas supersonically and 
producing a collisionless shock, indicating that the equations 
of strong shock physics can be applied to this 
region~\cite{Daly1990}, as discussed in Section \ref{StrongShockPhysics}. The source characteristics indicate that FRIIb radio galaxies can be used to determine and study the cosmological model that describes our universe, as described in Section \ref{CosmologicalStudies}. These studies allowed the total outflow lifetime to be empirically determined in a model-independent manner and independent of assumptions regarding minimum energy conditions, as described in Sections \ref{StrongShockPhysics}--\ref{TotalOutflowLifetime}.  The relationship between the total outflow lifetime and black hole mass is discussed in Section \ref{TvsM}. The luminosity in 
directed kinetic energy can be combined with the outflow lifetime to determine the total energy transported from the supermassive black hole 
via the jetted outflow over the source lifetime, 
which is studied relative to black hole mass in 
section \ref{TotalOutflowEnergy}. The results presented in section \ref{TotalOutflowEnergy} suggest a second method of studying the relationship between 
total outflow lifetime and black hole mass, which is presented in section \ref{TvsM-fM}. The results 
preseneted in section \ref{TotalOutflowEnergy} also suggest a 
relationship between the luminosity in directed kinetic energy (also referred to as the ''beam power'') and the total black hole 
mass, as described in Section \ref{LvsM-fM}. Finally, the total outflow lifetime is compared with the age of the universe at the redshift of each source to empirically determine the radio source selection function, which is presented in section \ref{TuoTo}. 

\subsection{FRIIb Radio Sources}
\label{FRIIb}

The FRIIb sources discussed and studied here are type 1 FRII sources, as defined in~\cite{LW84} and studied in detail \mbox{in 
\cite{LW84,AL87,LMS89,C91,Liu92,ODea2009}}, for example. These are the most 
powerful FRII sources and have 178 MHz radio powers greater than about 
$3 h^{-2} \times 10^{26} ~\rm{W ~Hz^{-1} sr^{-1}}$,  with radio powers obtained from radio flux densities using a standard LCDM cosmological model. Thus, FRIIb radio sources have radio powers that are greater than about \mbox{$8 \times 10^{27}$ W/Hz} at 178 MHz for a standard value of $h \simeq 0.7$. (Note that in work carried out a few decades ago, the 
standard cosmological model was an open, empty universe, and in that cosmological model, the 178 MHz power cut translates to about $h^{-2} \times 10^{27} ~\rm{W ~Hz^{-1} sr^{-1}}$.) They are selected from the 
3CRR survey~\cite{LRL83}. 

FRIIb radio sources include only 
the most powerful FRII sources. For comparison, the LOFAR sample of 
23,344 radio loud AGN includes fewer than about a hundred FRIIb sources, so only a small fraction---less than about 
0.4\%---of the those sources have 
\mbox{150 MHz} radio powers typical of FRIIb sources, that is, have radio powers greater than about $10^{28}$ W/Hz~\cite{Hetal19}. This means that general conclusions reached with the LOFAR study do not apply to FRIIb sources. The radio sources and results discussed here are largely consistent 
with those obtained by 
\cite{Blundell99}, including the conclusion that individual sources do not exhibit self-similar behavior. {Ref.}~\cite{Blundell99} states the following: ``We find that throughout the lifetime of an individual source its axial ratio steadily increases thus, for an individual source its expansion is not self-similar.'' 

In comparing the properties 
of the radio bridges of FRIIb sources with other studies, the results obtained by \citet{Daly2010} 
are quite instructive. In that work, each of 11 FRIIb radio galaxies   
was rotated into a horizontal position, and the properties of cross-sectional slices of the bridge were studied as a function of distance from the hotspot region, including  the cross-sectional radio surface brightness profile, the width of the bridge, the radio emissivity as a function of the position across each slice, the first and second moments of the slice surface brightness, the mean surface brightness, the minimum-energy magnetic field strength, and the mean pressure of the relativistic plasma.  
These studies indicate that for FRIIb sources, the region in the immediate vicinity of the radio hotspots produces most of the total radio power of the source, and the radio bridge has a roughly constant radio surface brightness, emissivity, and bridge width. The results are consistent with the description of the sources as summarized in Sections \ref{StrongShockPhysics} and  \ref{CosmologicalStudies} and are consistent with the results of similar studies by~\citet{LMS89}.  

The structure of the radio bridges of FRIIb radio galaxies indicate that the hotspot region, that is, each end of the ``cigar-shaped'' radio bridge, is moving into the ambient gas supersonically (e.g.,~\cite{AL87, LMS89, C91, Liu92, WDW97a, WDW97b, ODea2009, Daly2010}). Indeed, the shape of the radio bridge was used to infer the Mach number with which the 
radio bridge length increases, as well as the properties of the ambient gas~\cite{WDW97a, WDW97b}, as discussed in Section~\ref{StrongShockPhysics}. 

{Refs.} \cite{AL87,LMS89} conclude that the reason for radio bridge distortions exhibited by FRIIa sources (including four standard types and three complex types) is most likely the backflow of relativistic plasma in the radio bridge region. These authors conclude that backflow is not important for FRIIb sources and, thus, that the rate of growth of each source, referred to as the ``lobe 
propagation velocity,'' ($v_L$), as indicated by a radio spectral aging analysis, could be used to determine how rapidly the radio bridge length increases for FRIIb sources. Similar 
results have been obtained by other groups, such as \citet{C91}, who studied well-known FRIIb source Cygnus A.

The properties of the 31 radio galaxies discussed in Section \ref{CosmologicalStudies} are described by~\cite{DDetal08}. The sample includes the 11 radio galaxies presented by~\citet{Kharb2008}, with details on individual source properties presented by~\citet{ODea2009} and the 19 radio galaxies previously studied \mbox{in~\cite{GDW2000,DG02,PDMR2003}}. 
For some studies, source 3C427.1 was removed, as discussed by~\citet{Detal09}; these studies include 30 rather than 31 sources.

The determination of radio source parameters often depends upon the assumed underlying cosmological model. The effects of adopting different cosmological models is included in these studies, as discussed in Section \ref{CosmologicalStudies}. 

\subsection{The Strong Shock Method}
\label{StrongShockPhysics}

The radio bridge structure of FRIIb sources indicates that the hotspot region at the end of each side of the cigar-shaped radio bridge is moving into the ambient gas supersonically (e.g., \cite{LW84, AL87, LMS89, C91, Liu92, Blundell99, ODea2009}), and is producing a collisionless shock, which is mediated by the magnetic field that permeates the plasma (e.g.,~\cite{Daly1990}). This means that the equations of strong shock physics can be applied to the region 
just behind the hotspot region (i.e., in the direction toward the supermassive black hole) (e.g.,~\cite{BC89, Daly1990, RS91, C91,WDW97a,WDW97b,ODea2009}).   

Applying the equations of strong shock physics allows for empirical determinations of the rate at which energy is deposited into this region, referred to as the luminosity in directed kinetic energy ($L$); the ambient gas density ($n_a$); the 
ambient gas pressure ($P_a$); and the ambient gas temperature ($T_a$). In these empirical determinations, a term that describes the offset from minimum energy conditions of the relativistic plasma in the radio bridge region is included (e.g.,~\cite{C91, WDW97a, WDW97b, Belsole2007, Croston2005, ODea2009, Ineson2017}), as described below. 

For the work discussed here, the key quantities are $L$ (also referred to as the ``beam power'') and 
$n_a$. The luminosity in directed kinetic energy and the ambient gas density are obtained by applying the equations that are valid for a strong shock: $L \propto v_L P_L a_L^2$ and the ambient gas density is obtained with $n_a \propto P_L v_L^{-2}$, where $v_L$ is the rate at which the source length is increasing, sometimes referred to as the ``lobe propagation velocity''; $P_L$ is the pressure of the relativistic plasma in 
the post-shock region (i.e., just behind the hotspot region); and $a_L$ is the half-width of the radio bridge in the post-shock region 
(e.g.,~\cite{Daly1990, C91, D94, WDW97a, WDW97b, ODea2009, Daly2010}). 
Only radio galaxies were included in the study of~\citet{ODea2009} 
so as to minimize projection effects.  

Both $v_L$ and $P_L$ depend on offsets from minimum energy conditions in the post-shock region. Numerous observations and studies indicate that the relativistic plasma in the radio bridge region is offset from minimum energy conditions. The offset from minimum energy conditions is parameterized with $b \equiv B/B_{min}$, where $B$ is the magnetic field strength and $B_{min}$ is the minimum energy magnetic field strength obtained as described by~\citet{ODea2009}. Typical values of $b \simeq  0.3$ have been indicated by numerous empirical studies  
\cite{C91, Belsole2007, Croston2005, WDW97a, WDW97b, ODea2009, Ineson2017}). Results obtained with both $b \simeq 0.25$ and $b = 1$ are discussed and compared in
\cite{ODea2009,WDW97a,WDW97b}, for~example. 

A key result reported by \citet{ODea2009} is that the $b$  parameter cancels out in the empirical determination of $L$~\cite{Daly1990}, that is, the way that $b$ enters $P_L$ and $v_L$ cancel out since $v_L \propto b^{1.5}$ and $P_L \propto b^{-1.5}$ for $b \le 1$, as described in Section 4 of~\cite{ODea2009}. The beam powers ($L$) shown in {Figures} 28 and 29 of~\cite{ODea2009} 
as a function of core-hotspot separation for 31 FRIIb radio galaxies, assuming values of $b = 1$ and $b = 0.25$, respectively, illustrate that the luminosity in directed kinetic energy obtained 
with the strong shock method is not affected by offsets from minimum energy conditions.  

Thus, the empirically determined beam powers for the 31 FRIIb radio galaxies presented by~\cite{ODea2009} are independent of offsets from minimum 
energy conditions in the radio bridge and, therefore, are quite reliable. 
Another key result illustrated by {Figures} 
28 and 29~\cite{ODea2009} is that there is no indication of a relationship between ($L$) and the current radio-source size; this is also indicated by the results shown in {Tables 5 and 6} 
 of that work. This indicates that each source has a roughly constant value of $L$ during its lifetime. A roughly constant value of $L$ per source during the source lifetime is also indicated by the structural properties of the cigar-shaped radio bridges of FRIIb sources (e.g., \cite{LMS89,Daly2010}). 

A constant value of $v_L$ of a given source over its lifetime is also indicated by the properties of FRIIb radio sources, since there is no correlation between the lobe propagation velocity ($v_L$) and the current radio-source  size (see Tables 5 and 6 and {Figures} 22 and 23 of 
\cite{ODea2009}). This indicates that each source has a roughly constant 
value of $v_L$ during its lifetime. The lobe propagation velocity is also independent of redshift~\cite{ODea2009}. 

The $b$ parameter has an enormous impact on the ambient gas density, which scales as $n_a \propto b^{-4.5}$ (see Section 8.8.1 of~\cite{WDW97a} and Section 7.4 of~\cite{WDW97b}). This very strong dependence allowed for a constraint on the fractional dispersion of this parameter, and values of \mbox{$\delta(b)/b \leq 0.11$ for $b \simeq 0.25$}, and $\delta(b)/b \leq 0.09$ for $b \simeq 1$ were obtained~\cite{WDW97a}. A conservative bound on the source-to-source dispersion of the $b$ parameter of $\delta(b)/b \simeq 0.15$ 
was adopted by~\citet{WDW97a}. Thus, 
whatever the offsets from minimum energy conditions are within the radio bridge region, this offset must have a very small source-to-source dispersion for FRIIb \mbox{radio sources}. 

The ambient gas density as a function of distance from the host supermassive black hole indicates an empirically determined  composite density 
profile. The empirically determined composite profile is consistent with the host supermassive black hole  residing near the center of a gaseous environment, similar to clusters of galaxies at low redshift, such 
as the cluster that hosts Cygnus A~\cite{C91, WDW97a,WDW97b, Belsole2007, Croston2005}. 
{Ref.} \cite{WDW97a} concludes that FRIIb sources are located at the centers of cooling flow clusters of galaxies. This is consistent with results obtained by other groups for 
similar sources~\cite{C91,Belsole2007, Croston2005}. 

The shape of the radio bridge was used to measure the Mach number with which the forward shock front (in the vicinity of the radio hotspots) 
propagates into the ambient medium, which can be combined with $v_L$ to solve for the temperature of the ambient gas~\cite{WDW97a}. Combining the temperature and density 
obtained for a sample of 12 radio galaxies and 6 radio loud quasars with redshift between 0 and 
1.8, the authors of {ref.}~\cite{WDW97a} found that the ambient gas density and temperature values were in good agreement with independently determined values and 
indicated that the sources lie near the center of a ``cooling flow region,'' likely at the 
center of a cluster or proto-cluster of galaxies. Similar results have been obtained by other groups~\cite{C91, Belsole2007, Croston2005}. Thus, FRIIb radio sources are likely located at the center of cooling flow clusters 
or proto-clusters of galaxies. 

Detailed studies of FRIIb sources indicate that the region in the immediate vicinity of the radio 
hotspots produces most of the total radio power of the source, and the radio bridge has a roughly 
constant radio surface brightness, emissivity, and radio bridge width~\cite{LMS89, Daly2010}. The determination of each of these parameters depends upon the assumed underlying cosmological model. The effects of adopting different cosmological models was included in these studies. 

\subsection{Cosmological Studies}
\label{CosmologicalStudies}

Several properties of FRIIb radio galaxies indicated that they could provide a useful tool to study and constrain cosmological models 
\cite{D94, GDW2000, PDMR2003, ODea2009,  DDetal08, Detal09}. 
Detailed studies of FRIIb radio galaxies and radio loud quasars indicate that the radio bridge structures of radio galaxies are straight and regular, while those of radio loud quasars exhibit some  distortions~\cite{LMS89}. In addition, studies of FRIIb radio galaxies suggest that these sources likely 
lie close to the plane of the sky and, thus, are not strongly affected by 
projection effects~\cite{LMS89,B89,UP95,BP24}.   It was noted that the rate of growth of each source ($v_L$) for FRIIb radio galaxies is independent of source size and redshift, as discussed in Section \ref{StrongShockPhysics}. Thus, the empirical data are consistent with a 
constant values of $L$ and $v_L$ for a given FRIIb radio galaxy over its lifetime. For cosmological studies, the total luminosity in directed kinetic energy is used and is taken to be the weighted sum of $L$ from 
each side of each source. 

Images of powerful classical double radio sources, referred to here as FRIIb sources, are presented in~\cite{BP24}. Cyngus A, also known as 3C405, is an excellent example of an FRIIb source, and images of this source are presented in~\cite{Oetal25}.    

The 30 individual FRIIb radio galaxies discussed here were drawn from the 3CRR survey of radio galaxies~\cite{LRL83}, which is referred to as the 
``parent population,'' (see~\cite{DDetal08, Detal09} for a summary).  It was noted that the source sizes of the parent population 
(as inferred from the angular separation of the  radio hotspots on each side of each source) of the \mbox{70 galaxies} in the parent population first increase with redshift, then decrease with redshift over the redshift interval from about zero to two for all reasonable 
cosmological models 
(as summarized by~\cite{Detal09}; see 
also~\cite{GD98,GDW2000,DG02,PDMR2003}). The sizes of these intrinsic physical sources depend upon the assumed cosmological model, since the 
intrinsic size depends upon the angular size of the radio galaxy, 
the redshift of the galaxy, and the coordinate distance to that redshift (which is 
where the dependence on the cosmological model  enters). The effect of an assumed cosmological model on the sizes of physical sources was present in all reasonable 
cosmological models, though the magnitudes of the changes were different for different models. Cosmological models including a cosmological constant, space curvature, dark energy, and a rolling scalar field have been studied~\cite{DDetal08, Detal09}. 

As noted earlier, the velocity ($v_L$) with which an individual source is increasing in size was found to be independent of source size and independent of redshift~\cite{ODea2009}, where the size is measured from the location of the supermassive black hole to the 
hotspot on each side of each source. To reconcile the observed evolution of the mean source size of the parent population of radio galaxies with redshift with the lack of 
evolution of $v_L$, the ansatz that the total time the jetted outflow produced by the supermassive black hole ($T_*$) could be written as a power law of the beam power ($T_* \propto L^{-n}$), was proposed, with $n$ related to the radio source 
parameter ($\beta$, $n \equiv \beta/3$)~\cite{D94}. 
(Note that the beam power, $L$, is also referred to as luminosity in directed kinetic energy.) 
A characteristic source size ($D_* \propto v_L ~T_*$) was defined and compared with the mean source size of the parent population as a function of redshift~\cite{Detal09}. 

This comparison depends rather strongly on the underlying cosmological model, as described in Section 2.1 of~\cite{Detal09}. Of course, obtaining the source size from the observed angular size depends linearly on the coordinate distance 
($(a_0r)$), while the characteristic source size ($D_* \propto v_L L^{-n} \propto (a_0r)^{(4/7) - 2n}$; see Equation 3 and subsequent discussion presented in~\cite{Detal09}). 
Thus, requiring that values of $D_*$ track the mean size of the parent population with redshift has a dependence of $(a_0r)/[(a_0r)^{(4/7) - 2n}] \propto 
(a_0r)^{(3/7) + 2n}$. Requiring that the characteristic sizes of individual 
sources track that of the parent population as a function of redshift allows for determination of the 
best fit value of $n$ and 
determination of the underlying cosmological model 
\cite{D94, GD98, GDW2000, DDetal08, Detal09}. In each of these studies,  
the results consistently indicated that models that included a cosmological constant provided an excellent description of the data, and a flat, matter-dominated universe was clearly ruled out. 

The most recent constraint is obtained by fitting jointly to FRIIb radio galaxies and type Ia supernovae. The results indicate a value of $n = 0.50 \pm 0.05$~\cite{Detal09}, similar to and consistent with values obtained in earlier work. Thus, the ansatz expressed as $T_* \propto L^{-n}$ provides an excellent description 
of FRIIb sources, as discussed in more detail in Section~\ref{TotalOutflowLifetime}. Results obtained with FRIIb radio galaxies alone have consistently indicated a cosmological model that is in agreement with the currently accepted cosmological model for our \mbox{universe~\cite{D94, GD98, GDW2000, PDMR2003, DDetal08, Detal09}.} 
 
\subsection{The Total Lifetime of the Jetted Outflow}
\label{TotalOutflowLifetime}

The method to obtain and study the total outflow lifetime for FRIIb sources is summarized in Section \ref{CosmologicalStudies}; see Section 2.1 of ~\cite{Detal09}
for a more detailed description. 
The obtained jetted outflow lifetime ($T_*$)  is the total time that the AGN will actively be producing collimated dual jets that terminate in the radio hotspots. 

The empirically determined values of 
the luminosity in directed kinetic energy ($L$) and total jetted outflow lifetime ($T_*$) have no inputs related to the central black hole system, including no inputs related to the black hole or accretion disk (see Sections \ref{StrongShockPhysics} and \ref{CosmologicalStudies}). The key result for studies of the jetted outflow lifetime is that $n = 0.50 \pm 0.05$ (i.e., $\beta = 1.50 \pm 0.15$) and  
\begin{equation}
T_* \simeq 2.5 ~L_{46}^{-1/2} \times 10^7 ~\hbox{yr}
\label{eqTD}
\end{equation}
where $L_{46}$ is the total jetted outflow luminosity in directed kinetic energy, also referred to as the beam power, in units of $10^{46}$ erg/s. For the 100 FRII sources studied by~\citet{D19}, the mean value and standard deviation of the beam power was found to be $\rm{Log}(L) \simeq 
45.71 \pm 0.75$. This range of values of $L$ is nearly identical to the range of values for the sources studied by 
\citet{ODea2009}. The uncertainty of $T_*$ is dominated by the uncertainty of the luminosity in directed kinetic energy~\cite{ODea2009, D22}. Given that the uncertainty of $L_{46}$ is $\delta{\rm{Log(L_{46})}} \simeq 0.13$ (e.g.,~\cite{D19}), the uncertainty of the total outflow lifetime is 
$\delta{\rm{Log(T_{*})}} \simeq 0.065$. This indicates 
an uncertainty of $\delta T_*/T_* \simeq 0.15$, since 
$\delta Log(x) = (\delta x/x) ~[ln(10)]^{-1}$. It is certainly possible that the constant of proportionality of 2.5 could have a systematic offset by up to a factor of about 1.5, as discussed by~\citet{Detal09}. 

The empirically determined values for the luminosity in directed kinetic energy of individual sources applied for these studies are easily understood in the context of either the model proposed in \cite{BZ77} or in the context of the model of jet production~\cite{M99}. 
For example, Refs.~\cite{D09a, D09b, D19} showed that the empirically determined values of $L$ are consistent with predicted values for reasonable magnetic field strengths and black hole spin values, given the empirically determined values of black hole masses for these sources. Thus, even though the physical mechanism of 
jet production and the role of the magnetic field in launching the jets is quite different in these models, 
the equation relating $L$ to black hole mass, spin, and magnetic field strength are quite similar in both functional form and normalization, as discussed in detail by~\citet{D19}. Thus, it is unlikely that the intrinsic values of $L$ are significantly different from the empirically determined values of $L$. 

The fact that the total jetted outflow lifetime of FRIIb sources is well represented as a function of only 
the luminosity in directed kinetic energy, which must be controlled by processes associated with the 
central black hole, is one indication that empirical studies of these sources can be used to understand the properties of the black hole at the center of the black hole system. 

\subsection{The Relationship Between the Total Lifetime of the Jetted  Outflow and Black Hole Mass}
\label{TvsM}

The total lifetime of the jetted outflow given by Equation (1) is obtained independent of the properties of the central region of the radio source. Thus, it is obtained independent of the black hole mass and accretion disk properties. The outflow lifetime can be compared with the mass of the host supermassive black hole to obtain the relationship between total outflow lifetime and black hole mass. 

Of the original 31 radio galaxies studied by \citet{ODea2009}, 19 had published supermassive black hole mass values, and these sources were studied by~\citet{D09a}. To increase the number of sources, the relationship between $L$ and the 178 MHz radio power ($P$) for the sample of 31 sources mentioned above was obtained~\cite{Detal12}. The empirically determined relationship, 
$\rm{Log}(L) \propto (0.84 \pm 0.14) \rm{Log}(P)$, was shown to be consistent with expectations based on 
the application of the strong shock method, given that the radio power is one of the factors that enters into the empirical determination of $L$. Thus, it is unlikely that the empirically determined relationship between $L$ and $P$ results from a Malmquist bias due to a common dependence of these quantities on  distance. To test for a Malmquist bias, the standard method of applying a partial correlation analysis was carried out using the method and code of~\cite{AS96}, which is described in detail by~\citet{T21}. The method is based on the application of Kendall's partial $\tau$ test, which numerically tests for a correlation between two parameters after removing the effect of a third parameter on their correlation. Following standard practice, redshift is used as a proxy for distance, applying the method to the 31 sources studied by~\citet{Detal12}, with the results indicating that the correlation between $L$ and $P_{178}$ obtained by 
\citet{Detal12} is significant at greater than about 97\% confidence after removing any common dependence of these parameters on distance. 

Applying the relationship between $L$ and $P_{178}$ allowed the sample size to be significantly expanded, and the sample of 97 FRIIb radio sources with values of both $L$ and the black hole mas ($M$), as presented and discussed by~\citet{D16}, are studied here. The black hole masses are obtained from~\cite{McLure04, McLure06}, and the source types are from~\cite{LRL83}.
Values of these quantities, plus values for three additional sources, 
are listed by~\citet{D19}. Black hole masses and bolometric luminosities were available for 17 of the radio galaxies studied by~\citet{ODea2009}, and these sources are included in the samples studied in~\cite{D16,D19}. For the remaining sources, Equation (1) is applied to convert $L$ to $T_*$.  

\begin{figure}[H] %
\includegraphics[width=\columnwidth]{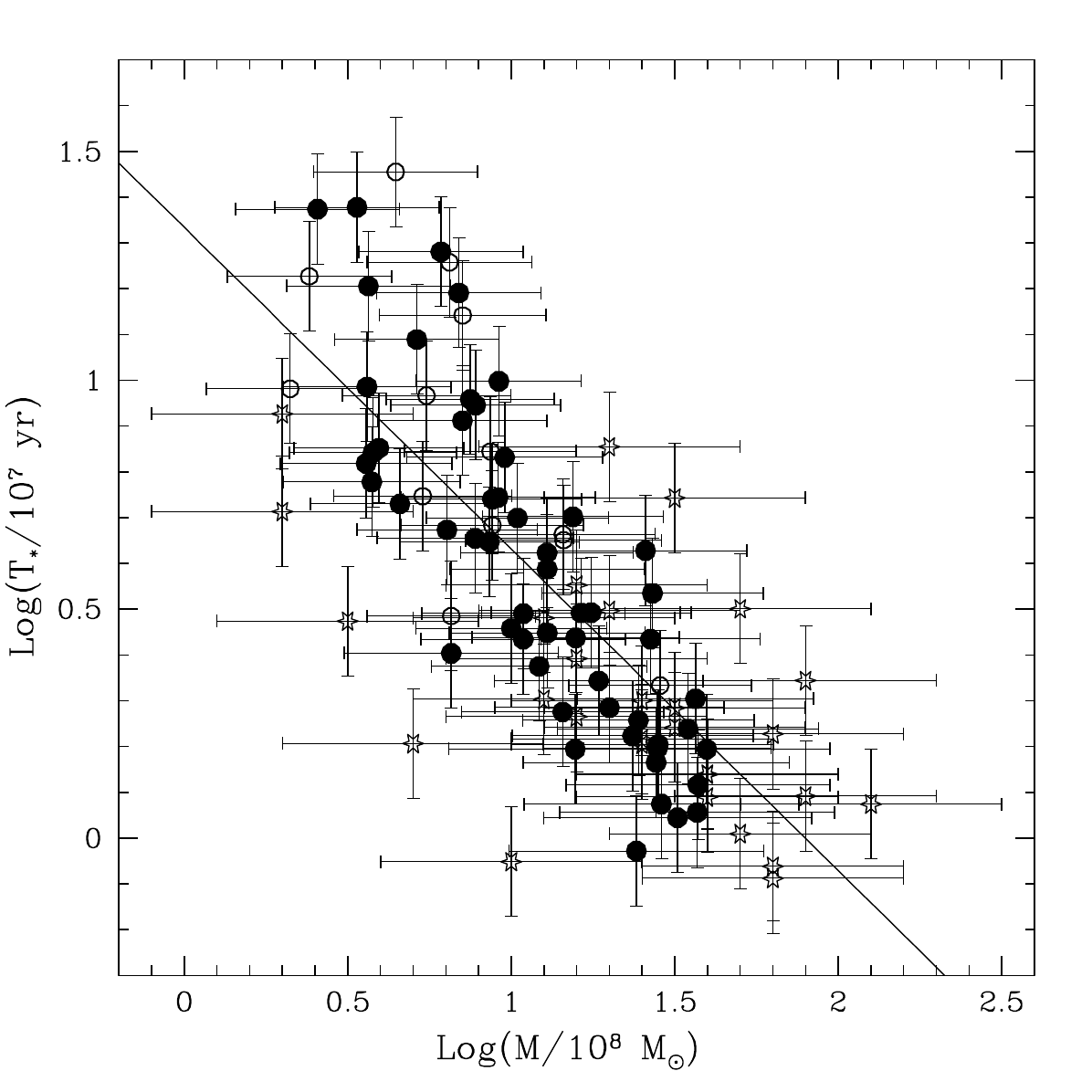}
    \caption{The total
 lifetime of the jetted outflow (in units of $10^7$ yr) is shown versus black hole mass (in units of $10^8 M_{\odot}$) for the sources discussed in \ref{TvsM}. Included are samples of 55 high-excitation radio galaxies (solid circles, HEG), 29 radio loud quasars (open stars, Q), and 13 low-excitation radio galaxies (open circles, LEG). These designations are based on the nuclear spectroscopic properties of the sources. For $\rm{Log}(T_7) ~vs~ \rm{Log}(M_8)$, as illustrated in the figure, the following (slope; y-intercept) pairs are obtained:  
    55 HEG ($-0.92 \pm 0.08$; $1.57 \pm 0.09)$; 
    29 RLQ ($-0.33 \pm 0.09$; $0.75 \pm 0.13)$; 
    13 LEG ($-0.72 \pm 0.24$; $1.5 \pm 0.2)$; 
    the full sample of 97 sources ($-0.70 \pm 0.06$; $1.33 \pm 0.07)$. The fit to the full sample is shown on the figure. All fits are unweighted, and an uncertainty of $\delta \rm{Log}(T_*) \simeq 0.12$ applies to values of $T_*$ obtained as described in section \ref{TvsM}. Results obtained with an independent approach indicate that $\rm{Log}(T_*/10^7 \rm{yr}) \simeq - \rm{Log}(M_8) + 1.5$,  which is consistent with the results presented here for HEG 
    (see Sections \ref{TotalOutflowEnergy} and \ref{TvsM-fM}). 
    }
    \label{fig1:Fig1}
\end{figure}

Since the empirically determined values of $T_*$ and $M$ are obtained independently, it is important to test for a Malmquist bias, that is, to test whether the relationship illustrated by Figure 
 \ref{fig1:Fig1} is intrinsic to the sources or is a result of a common dependence of $T_*$ (or $L$) and $M$ on distance. Applying the technique of~\citet{AS96} described above, the results indicate that the correlation between $T_*$ and $M$ for the full sample of 97 sources is significant at greater than 99.98\% confidence after removing any common dependence of these parameters on distance. The result for the sample of 55 HEG alone is significant at greater than 98.9\% confidence after taking into account any common dependence on distance, and for the sample of 29 RLQ, this confidence is about 94.5\%. However, for the sample of \mbox{13 LEG}, the confidence is only 82\%, which likely results from  the small sample size. Thus, the correlations between the total outflow lifetime and black hole mass are reliable for the full sample of 97 sources and the sub-samples of 55 HEG and  29 RLQ. Figure \ref{fig1:Fig1} illustrates the relationship between $T_*$ and $M$ obtained for the sample of 97 sources studied; best fit parameters for the full sample and each subsample and  are listed in the figure caption.  

\vspace{-3pt}

It should be noted that in the event that the outflow lifetime is set by a time scale related to a binary black hole system, the mass discussed here is that of the primary supermassive black hole, and the jetted outflow is taken to be anchored to the primary black hole, as discussed in Section \ref{BH Coalescence}. The relationships obtained in the following sections are fully consistent with the outflow method of measuring the black hole spin and accretion disk properties of the primary supermassive black hole~\cite{D19}. 

\subsection{Total Energy Deposited in Hotspot Region Relative to Black Hole Mass}
\label{TotalOutflowEnergy}

The total energy removed from the central black hole system and deposited into the hotspot region over the lifetime of the source is $E_T = L T_* \propto L^{1/2}$, where $L$ is the weighted sum of the beam power (i.e., luminosity in directed kinetic energy) 
from each side of the dual outflow. This total energy ($E_T$) was studied relative to the black hole mass ($M$) for a sample of 19 FRIIb radio galaxies for which both $L$, determined using the strong shock method (described in Section  \ref{StrongShockPhysics}) and $M$ were available~\cite{D09a}. The distribution of values of $f_M \equiv E_T/(Mc^2)$ was found to have a  very small dispersion. This was not the case for the sample of FRI ``cavity'' sources, which was found to have a very broad distribution of values, ranging from about $10^{-6}$ to about $10^{-2}$~\cite{D09a}. 

Of the 19 FRIIb radio galaxies included in the studies of~\cite{D09a}, 17 (all but 3C 55 and3C405, also known as Cyngus A) had sufficient information to be included in the studies of~\cite{D22}; all of these sources are HEG galaxies based on their nuclear spectroscopic properties. For these 17 sources, the mean value of $\rm{Log(f_M)} \simeq -2.49 \pm 0.33$ indicates the mean value of the total energy carried by the jetted outflow over the total outflow lifetime relative to the supermassive black hole mass of 
\begin{equation}
f_M \equiv E_T/(M c^2) \simeq (3.2 \pm 2.4) \times 10^{-3}~. 
\end{equation}

Each of the 17 studied sources has a total luminosity in directed kinetic energy obtained with the strong shock method and; thus, they are independent of assumptions, including assumptions regarding offsets from minimum energy conditions, as reviewed in Section \ref{StrongShockPhysics}, and, as such, are \mbox{quite reliable}. 

To study $f_M$ for larger samples of HEG FRIIb sources, the relationship between $L$ and the 178 MHz radio 
power was obtained as described in Section \ref{TvsM} and applied, which 
led to a sample of 55 HEG FRIIb radio galaxies with values for both $L$ and $M$~\cite{D16}. The total outflow energy ($E_T = L T_*$) was obtained and studied relative to the black hole mass ($M$) for this sample by~\citet{D22}, which indicated a value of 
\begin{equation}
f_M = E_T/(M c^2) \simeq (3.5 \pm 1.5) \times 10^{-3}~. 
\end{equation}

Similar values with a slightly larger uncertainty are obtained for other source types; see Table 3, column 9 of~\cite{D22}. The sample of 55 HEG radio galaxies includes the 17 sources for which $L$ and $E_T$ are obtained as described above and in Sections \ref{CosmologicalStudies} and~\ref{TotalOutflowLifetime}.

It is remarkable that even though the FRIIb sources discussed here 
have a broad range of values for each of the parameters ($\rm{L}$, $\rm{E_T}$, $\rm{M}$, and redshift (z)), they have a very small range of values of $\rm{E_T/(M c^2})$, as illustrated in Figure 19 of~\cite{D22}. In addition, the quantity $f_M$ has no evolution with redshift, as indicated by the slope listed in column (9) of {Table 3} 
 for the full sample of the 100 studied sources and as illustrated in Figure 20 of~\cite{D22}. 

The study reported in~\cite{D22} included a comparison of values of $f_M$ obtained for 100 supermassive black holes with different nuclear spectroscopic properties. In that work, high-excitation galaxies (HEG), low-excitation galaxies (LEG), quasars (Qs), and weak radio galaxies (W) were studied. There is no statistically significant difference in the mean  values of $f_M$ for sources with 
different nuclear spectroscopic properties (see column 9 of Table 3 in~\cite{D22}). Thus, there is no correlation between the nuclear spectroscopic properties of the gas in the immediate vicinity of the supermassive black hole and the total energy extracted from the supermassive black hole system relative to the black hole's mass, though HEG sources exhibit a smaller dispersion relative to other source types, even after accounting for the larger number of sources in the HEG sample. 

This is a second indication that a process directly related to the supermassive black hole is controlling the lifetime of the dual collimated outflow, with the first indication being Equation (1).

The facts that $f_M = E_T/(Mc^2)$ appears to be a constant and that $E_T = L T_* \propto L^{1/2} \propto T_*^{-1}$ suggest that $T_* \propto M^{-1}$ or $L \propto M^2$---or both. This suggests that the relationship between the total lifetime of the jetted outflow and supermassive black hole mass could be obtained from $T_* \propto (f_M M)^{-1}$, as discussed in Section \ref{TvsM-fM}. It also suggests that the relationship between luminosity in directed kinetic energy and black hole mass could be obtained from $L \propto (f_M M)^2$, as discussed in Section \ref{LvsM-fM}. 

\subsection{The Relationship Between Jetted Outflow Lifetime and Black Hole Mass Indicated by $f_M$}
\label{TvsM-fM}

A direct measurement of the relationship between the total outflow lifetime and black hole mass is discussed in Section \ref{TotalOutflowEnergy}. It is suggested there that another indication of the relationship between the total lifetime of the jetted outflow and the black hole mass can be obtained by noting that $f_M$ has a small dispersion. 

The relationship between the lifetime for which the large-scale jetted outflow persists and the luminosity in directed kinetic energy indicated by Equation (1) allows Equation (2) or (3) to be re-written as 
$T \propto f_M^{-1} M^{-1}$. Substituting in the constants, Equation (2) or (3) indicates that 
\begin{equation}
T_7 \simeq 0.11~ f_M^{-1}~ M_8^{-1}~,
\end{equation}
where $T_7$ is the jetted outflow lifetime in units of $10^7$ yrs and $M_8$ is the black hole mass in units of $10^8 M_{\odot}$. 

Given that the $f_M$ obtained for HEG radio galaxies has a quite narrow distribution, as indicated by Equation (3), substituting in the value of $f_M$ from this equation suggests that 
\begin{equation}
T_7 \simeq 30 ~M_8^{-1},
\end{equation} 
where the coefficient has an uncertainty of about 40 \%. Values of  
$f_{M}$ for the 55 HEG, 29 Q, and 13 LEG, plus 3 W sources , are listed in column 9 of Table 3 of~\cite{D22}, and the mean values of $f_M$ for these samples are all quite similar, though the dispersion for the HEG is significantly smaller than for the other samples or for the sample of all 100 sources studied in that work. The sources studied by~\citet{D22} are the same sources illustrated in Figure \ref{fig1:Fig1}. 

\subsection{The Relationship Between Beam Power and Black Hole Mass Indicated by $f_M$}
\label{LvsM-fM}

Noting that the total outflow energy ($E_T$) is equal to $LT_*$, it is easy to see that \mbox{Equations (1) and  (2)} can be combined to obtain an empirically determined relationship between the luminosity in directed kinetic energy and the supermassive black hole mass, as discussed in \ref{TotalOutflowEnergy}. If so, this 
could be applied to FRIIb radio sources to obtain an estimate of the luminosity in directed kinetic 
energy based on the black hole mass.
Combining Equations (1) and (2), the empirically determined relationship is 
\begin{equation}
L_{46} \sim 0.65 ~(M_9)^2, 
\end{equation}
where $M_9 \equiv M/(10^9 M_{\odot})$. 
Of course, this relationship can 
only be applied to FRIIb radio sources, as described in Section \ref{FRIIb}, and is likely 
to have substantial scatter. 

\subsection{Jetted Outflow Lifetime Relative to the Age of the Universe}
\label{TuoTo}

The total outflow lifetimes obtained for the studied FRIIb sources, given by \mbox{Equation (1)}, range from about $10^7$ to a few $\times 10^8$  years. These lifetimes are long compared with human timescales, and the properties of an extended FRIIb source do not change over observed or human timescales. 
However, the lifetime is short compared with the age of the universe at the redshift ($z$) of the source. Thus, the probability of detecting one particular FRIIb source from the full population of such sources at that redshift is given by the ratio of the source lifetime to the age of the universe at that redshift ($T_{outflow}/T_{universe})$, where $T_{outflow}$ is given by Equation (1). This is referred to as the ``selection function.'' 

\vspace{-3pt}

\begin{figure}[H] %
\includegraphics[width=\columnwidth]{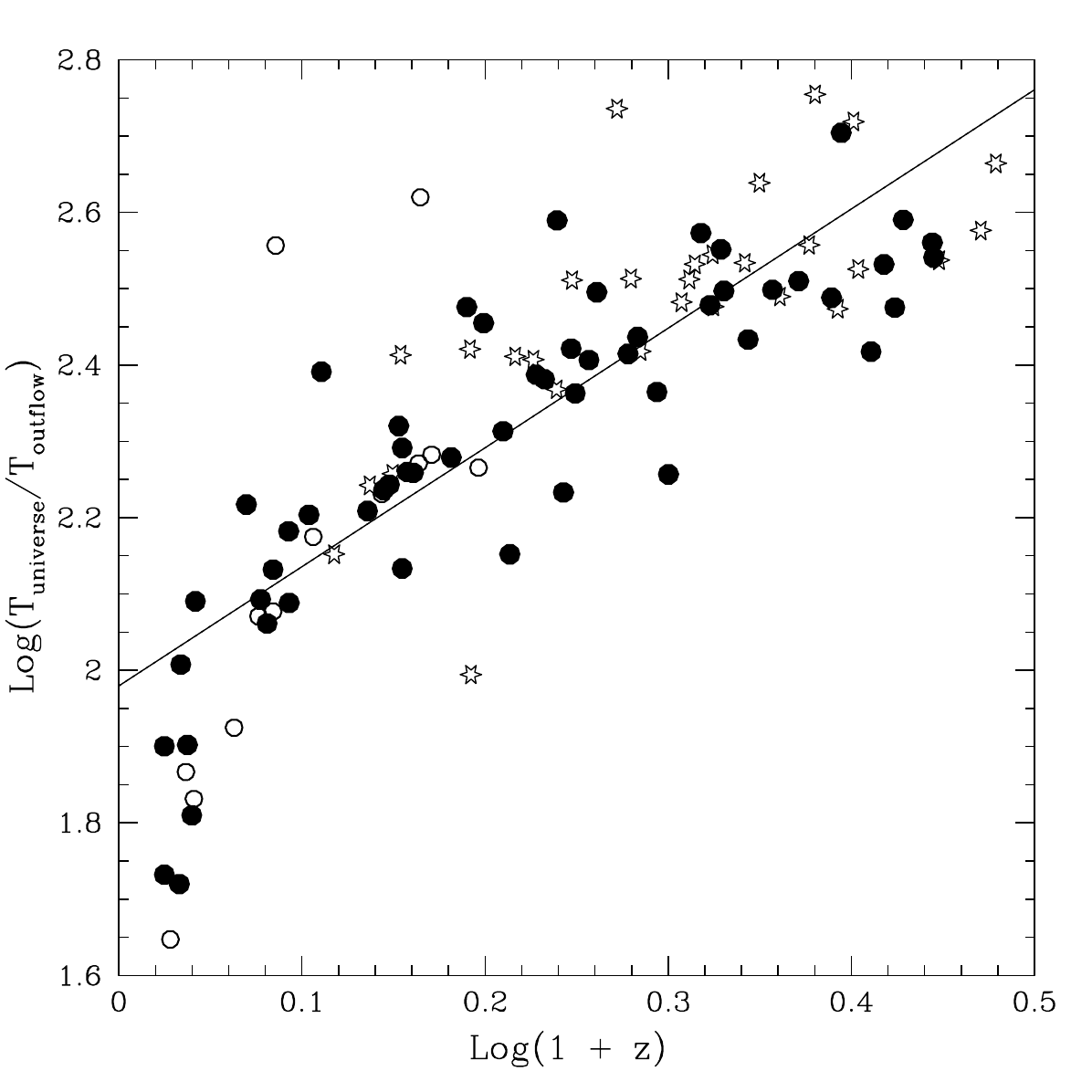}
\caption{The ratio 
 of the age of the universe at the source redshift ($T_{universe}$) to  total outflow lifetime ($T_*$; also referred to as $T_{outflow}$) is shown as a function of $(1+z)$, where $z$ is the source redshift. 
 The symbols and source types are as in Fig. 1. 
 As discussed in Section~\ref{TuoTo}, the quantity 
    of $T_{outflow}/T_{universe}$ is a measure of the incompleteness of the sample. Thus, its inverse  ($T_{universe}/T_{outflow}$) provides an empirical estimate the total of the number of FRIIb sources relative to the observed number at a given redshift. 
   The uncertainty of $\rm{Log}(T_{universe}/T_*)$ is about 0.12 since it is dominated by the uncertainty of $\rm{Log}(T_*)$ (see Fig. 1 caption). 
    The unweighted best fit line for the full population of 97 sources, $\rm{Log}(T_{universe}/T_{*}) \simeq \rm{Log}(1.56 \pm 0.11) ~\rm{Log}(1+z) + (1.98 \pm 0.03)$, is shown on the figure. 
    The best fit (slope;  y-intercept) pairs obtained for different source types are ($1.50 \pm 0.13$; $1.98 \pm 0.03)$ for 55~HEG; ($1.20 \pm 0.22$; $2.12 \pm 0.07)$ for 29 RLQ; ($3.60 \pm 0.99$; $1.76 \pm 0.12)$ for 13 LEG.
     All fits are unweighted. 
    }
    \label{fig1:Fig2}
\end{figure}

The sources that comprise the parent population are expected to have properties, including jetted outflows, 
outflow lifetimes, and black hole masses, that are similar to 
those measured for the detected sources. To obtain the number density of sources that comprise the parent population at a given redshift, the number density of observed FRIIb sources at that redshift should be multiplied by the ratio of $(T_{universe}/T_{outflow})$. This provides an empirical estimate of the number density of the full population of FRIIb sources at that redshift. The results illustrated in Figure \ref{fig1:Fig2} indicate that 
\begin{equation}
\rm{Log}(T_{universe}/T_{*}) \simeq (1.56 \pm 0.11) ~\rm{Log}(1+z) + (1.98 \pm 0.03)
\end{equation}
for the studied sample of 97 FRIIb sources; fits to subsamples are listed in the caption of Figure \ref{fig1:Fig2}.

\section{Connections to Black Hole Physics}
\label{Connections}

Empirical results obtained for FRIIb sources indicate that studies of these sources can be used to understand the physical properties and processes associated with supermassive black holes that produce these types of outflows, as described in Section 2, with a focus on results presented in 
Sections\ref{TotalOutflowLifetime}--\ref{TuoTo}. Three possible causes for the well-defined total outflow lifetime (see Equation (1)), fixed fraction of black hole mass 
extracted during this lifetime (see Equations (2) and (3)), 
and the implications of these results (see \mbox{Equations (4)--(7)} and {Figures} \ref{fig1:Fig1} and \ref{fig1:Fig2}) are discussed here. 
These particular properties only apply to FRIIb sources and do not apply to other types of jetted outflows. 

First, the jetted outflow could terminate or change character due to a change of the properties of the black hole, such as a change in the spin mass energy available to be extracted relative to the irreducible and total black hole mass, as discussed in Section~\ref{Changes-in-BH-Properties}. 
Second, if the outflow is produced during the coalescence of two black holes, the outflow could be significantly modified or terminated when the coalescence is complete or has reached a particular stage, as discussed in Section~\ref{BH Coalescence}. Third, the jetted outflow could be terminated or significantly changed due to a change in the accretion disk properties so that the disk properties required to produce the outflow are no longer present, as discussed in Section \ref{Changes-in-AD-Properties}. 

A combination of processes could also be the cause of the termination of the jetted outflow that produces the FRIIb radio source. For example, the accretion disk properties required 
for the production of the jetted outflow that feeds the FRIIb source could be significantly modified due to 
presence of a second supermassive black hole or due to the coalescence of two supermassive black holes. This, in turn, would be related to the production of the gravitational wave background. In this scenario, the primary supermassive black hole would produce the FRIIb radio source, and the jetted outflow from the primary would turn off or be significantly modified so as to no longer be categorized as an FRIIb source when the accretion disk properties are sufficiently disturbed by the 
presence of or coalescence with the secondary supermassive black hole. 

\subsection{Changes in the Black Hole Properties Such as the Irreducible Black Hole Mass}
\label{Changes-in-BH-Properties}

Changes in black hole properties that could impact jet production include changes in the spin mass energy available for extraction and/or  changes in the irreducible black hole mass relative to the total black hole mass~\cite{Ketal21,Setal21,Tetal21,RR23,RR24,R25}. If the jetted outflow is ultimately powered by black hole spin mass-energy extraction, this extraction must be accompanied by an increase in the black hole irreducible mass~\cite{Ketal21,Setal21,Tetal21,RR23,RR24,R25}.  Different physical processes that lead to spin mass-energy extraction simultaneously increase the black hole irreducible mass by different amounts. The efficiency of the spin mass-energy extraction should be parameterized as the ratio of the extracted spin mass energy relative to that converted to black hole irreducible mass~\cite{R25}. Some processes, such as those involving pair production, lead to rather low efficiency values of a few percent~\cite{P69,R25}, while others---especially those that involve electric and magnetic fields---can be highly efficient~\cite{R25}. The efficiencies are obtained through a series of iterative computations carried out for various sets of initial conditions. The key point is that the extraction of spin mass energy from a rotating black hole via electric and magnetic fields can be highly efficient. Another key point is that such a process can terminate long before most of the spin mass energy has been exhausted, so some 
of the spin mass energy is extracted, while some is converted to irreducible mass, and when the process terminates, the black hole can still have a significant amount of spin mass energy. 

However, interactions involving magnetic fields do not always have a relatively high efficiency. For example, the interaction of a large-scale magnetic field that is constant in magnitude and direction will only convert black hole rotational energy to black hole irreducible mass and will not lead to the extraction of any of the spin 
mass energy~\cite{P72,KL77} and, thus, has an efficiency of zero. Thus, the efficiency must be computed by accounting for the details of each specific process~\cite{R25}. 

Although the value of the irreducible black mass can be obtained empirically for astrophysical black holes~\cite{D22,DDO24}, there is no direct way to measure the change in the irreducible black hole mass during an outflow event. Since it is challenging to empirically determine the mass energy converted to irreducible black hole mass during an outflow event for astrophysical 
black holes, the efficiency is sometimes defined as the ratio of the extracted spin mass energy relative to the total black hole mass or relative to the spin mass energy available for extraction, which can be empirically determined (see \mbox{Equations (2) and (3)} in Section~\ref{TotalOutflowEnergy}). Values were obtained and studied by \citet{D22} for a sample of 100 FRIIb sources, 
and values were also been obtained for Sgr A* and M87*~\cite{DDO24}. 
 
Interpreting the observed ratio of extracted energy relative 
to total black hole mass of about 
$f_M = E_T/(M c^2) \simeq (3.4 \pm 2.1) \times 10^{-3}$  or 
$f_M = (3.5 \pm 1.5) \times 10^{-3}$ 
(see Section \ref{TotalOutflowEnergy}) 
as empirically determined efficiency factors would indicate an efficiency of extracted energy relative to black hole mass of about 0.3\%. However, this is assuming that the process ends when all of the available spin mass energy has either been extracted or converted to black hole irreducible mass, and the results of \cite{R25} suggest that processes may terminate well before this occurs. It remains to be determined which models of spin mass-energy conversion to black hole irreducible mass and spin mass-energy 
extraction would lead to the empirically determined properties of FRIIb sources described in Sections \ref{TotalOutflowLifetime}--\ref{LvsM-fM}. 

\subsection{Dual Black Hole Coalescence}
\label{BH Coalescence}

\textls[-15]{Recently, {Refs}. \cite{KHF25,Nanograv23b}, for example, 
showed that the nanohertz gravitational wave \mbox{background~\cite{Nanograv23a,EPTA23, Reardon23,Xu23}}} could be produced by quasars if most quasars are associated with coalescing supermassive black holes. The supermassive black hole parameters required to account for the nanohertz gravitational wave background have significant overlap with the properties of the FRIIb sources discussed in Section \ref{TheRadioSources}. For example, 
Figure 3 of~\cite{KHF25} shows that 
an excellent fit to the nanohertz gravitational wave background is obtained for a coalescence timescale of about $2.7 \times 10^{7}$ {yr. } 
This is remarkably similar to the total outflow lifetime measured for FRIIb sources, as indicated by Equation (1). For example, the mean value of the luminosity in directed kinetic energy for the 100 sources discussed here is $L_{46} \simeq 0.5$~\cite{D19}, indicating a total outflow lifetime of about $3.7 \times 10^7$ years for a typical FRIIb source.  

Thus, it seems possible that the empirically determined outflow lifetime for FRIIb sources 
could be set by the coalescence timescale for 
two supermassive black holes. In this scenario, the dual-jetted outflow would be produced by the primary supermassive black hole as two supermassive black holes are coalescing. The jetted outflow would 
cease, or the radio source type would be significantly modified so that the source is no longer categorized as an FRIIb source when the secondary disturbs the accretion disk properties of the primary sufficiently to substantially affect the jetted outflow or when the coalescence is complete. Independent empirical  support for coalescing supermassive black holes is discussed by~\citet{GB08}. 

To investigate a possible connection between the properties of the black holes discussed here and the production of the nanohertz gravitational wave background, it is instructive to compare the ranges of supermassive black hole mass, the Eddington ratio, and the outflow 
lifetimes for the 97 FRIIb sources studied here with those required to produce the nanohertz gravitational wave background. The supermassive 
black hole masses of the 100 studied FRII sources fall in the range of $(10^8 - 10^{10}) M_{\odot}$~\cite{D16, D19}. This is identical to the range of black hole masses discussed by~\citet{KHF25}. The 
bolometric luminosity relative to the Eddington luminosity for the 100 FRIIb sources discussed here falls within the range of about $\rm{Log}(f_{Edd}) 
\simeq (-1.65 ~\rm{to} -0.11)$~\cite{D16,D19}, which is similar to the range shown in Figure 4 of~\cite{KHF25} of about ($-$2 to 0). The redshift range of the FRIIb sources studied here is fairly evenly distributed between zero and two, which is slightly lower than that discussed by~\citet{KHF25} of about one \mbox{to three.}

The supermassive black hole parameters required to produce a nanohertz gravitational wave background are very similar to empirically determined 
values for FRIIb sources. In this interpretation, the empirically determined relationship between the
total outflow lifetime given by Equation (1) and the luminosity in directed kinetic energy  
would be identified as the 
coalescence timescale. In this case, the jetted outflow leading to the powerful classical double radio source would terminate when the two supermassive black holes have coalesced into one supermassive black hole or when non-gravitational forces between the black holes begin to dominate the black hole--black hole interaction. At this time, the newly formed black hole would have a total mass and spin value determined by a combination of the orbital, rotational, and mass properties of the black holes prior \mbox{to coalescence.} 

One clue as to whether this is a realistic possibility is to consider whether the process has the well-defined relationship between the outflow lifetime and black hole mass discussed in Sections \ref{TvsM}--\ref{TvsM-fM}. 
Another clue is whether the measured amplitude of the gravitational wave background matches that predicted after the number density of sources contributing to the background is computed using the inverse of the selection function shown in Figure \ref{fig1:Fig2}, as discussed in Section \ref{TuoTo}. It is also likely that the black holes could have  interactions that involve 
a combination of gravitational and other forces. 

To summarize, the timescale that seems to provide a reasonable fit to the observed gravitational wave background 
is similar to the empirically determined outflow lifetime indicated by Equation (1).  
Thus, a second possibility is that the 
outflow lifetime is related to the coalescence of two black holes. This is an intriguing possibility. 
The supermassive black hole parameters 
required to produce the nanohertz gravitational wave background are remarkably similar to the properties of black holes that produce FRIIb 
radio sources. Especially interesting is the fact that the coalescence timescale required to produce the observed nanohertz background is quite similar to that obtained with completely independent methods for FRIIb radio sources. All other FRIIb source parameters are quite similar to those required for supermassive black hole coalescence to produce the gravitational wave background, including the the bolometric accretion disk luminosity in Eddington units, the black hole mass range, and the redshift range. Independent empirical 
support for coalescing supermassive black holes 
is discussed by~\cite{GB08}. 

This interpretation   
provides insight into the physical origin of the empirically determined relationship described by eq. (1) in 
section 
\ref{TotalOutflowLifetime}. This relationship would arise from the combination of equations $L \propto M^2$ and $T_* \propto M^{-1}$, as discussed in section \ref{TotalOutflowEnergy}. 
A total outflow lifetime 
$T_* \propto M^{-1}$, with a constant of proportionality 
that agrees the 
empirically determined value, listed in 
section \ref{TvsM-fM},  
is consistent with that expected for coalescing black holes that are producing gravitational waves (e.g. \citep{KHF25,Nanograv23b}). 
And, 
the relationship $L \propto M^2$, with the empirically determined 
constant of proportionality listed in section 
\ref{LvsM-fM},  
is consistent with that expected for a spin-powered outflow \citep{D19,BZ77,M99,D11}. 

\subsection{Changes of Accretion Disk Properties}
\label{Changes-in-AD-Properties}

FRIIb sources are known to lie at the centers of cooling flow regions, as discussed in Section~\ref{StrongShockPhysics}. It is possible that jetted outflows that produce an FRIIb radio source could turn off or change characteristics so as to drop out of the FRIIb category due to changes in the mass accretion rate. The idea is that as energy is deposited into the circumgalactic medium by the jetted outflow, the 
ambient gas is heated, decreasing the ambient gas density and leading to a decrease in the mass accretion rate (e.g.,~\cite{MH07,D09a,DV22, HB23}).  
In this scenario, when the mass accretion rate decreases to some specific value or by some specific amount, the strength of the accretion disk's magnetic field decreases, and the outflow is terminated or the properties of the outflow are modified so that the extended radio source would no longer be classified as an FRIIb source. 

For this to be able to explain the FRIIb lifetime indicated by Equation (1), which leads to Equations  
(4)--(6) when combined with Equation (2) or Equation (3), would require that when a total of about $0.3 \%$ of the mass of the supermassive black hole is deposited into the hotspot region of the FRIIb radio source, 
the mass accretion be effectively shut off, independent of the size of the radio source and independent of the value of the luminosity in directed kinetic energy. 

Values of $f_M$ were obtained for supermassive black hole systems with different nuclear spectroscopic types, including high-excitation galaxies, low-excitation galaxies, and quasars~\cite{D22}. As discussed in Section \ref{TotalOutflowEnergy}, there is no correlation between the nuclear spectroscopic type, which depends on the properties of the gas in the immediate vicinity of the supermassive black hole, and 
the total energy extracted from the supermassive black hole system relative to black hole mass. Such a correlation would be expected if the mass accretion rate is affected by the 
energy input to the hotspot region by the jetted outflow, so this suggests that the outflow does not affect the mass accretion rate. 

In addition, dimensionless mass accretion rates for the 100 FRIIb sources discussed here were obtained and compared for sources with different nuclear spectroscopic types~\cite{D21}. The results indicate that there are mild variations, with quasars having the largest dimensionless mass accretion rates, followed by high-excitation galaxies; low-excitation galaxies have the lowest dimensionless mass accretion 
rates. Despite these variations, as noted above, there is no variation in $f_M$ with nuclear spectroscopic type. Thus, it seems unlikely that the source properties described in Sections~\ref{TotalOutflowLifetime} and \ref{TotalOutflowEnergy} could be explained as a result of changes related to the mass accretion rate. 

However, to explore this possibility further, it is interesting to consider whether feedback of this type is a factor for other types of radio sources. \citet{MH07} studied 15 ``cavity'' radio sources in clusters of galaxies powered primarily by FRI-type radio sources and identified a relationship between the luminosity in directed kinetic energy ($L_{dKE}$) input by the FRI radio source to the cavity, the black hole mass, and accretion disk bolometric luminosity 
($L_{bol}$) of the host supermassive black hole, with the black hole mass 
parameterized by the Eddington luminosity ($L_{Edd}$). The relationship reported by~\citet{MH07} has the form of
\begin{equation}
\rm{Log}(L_{dKE}/L_{Edd}) = A ~\rm{Log}(L_{bol}/L_{Edd}) +B,
\label{FL}
\end{equation}
\noindent with $A = (0.49 \pm 0.07)$ 
and $B = (-0.78 \pm 0.36)$. {Ref}. \cite{MH07} found no statistically significant correlation between Bondi power and bolometric accretion disk luminosity for these sources and concluded that a significant fraction of the Bondi power must be removed from the black hole system by the jetted outflow. 

In a related study, the energy input to each cavity radio source relative to the supermassive black hole mass was studied by~\citet{D09a} for a similar sample of 29 cavity radio sources with \mbox{1 source} that overlaps with those studied by~\citet{MH07}. These sources exhibit a very  broad range of values of $E_T/(M c^2)$. The values measured span the range of about $(5.2 \pm 2.8) \times 10^{-7}$ for M87 to about $(1.8 \pm 1.6) \times 10^{-2}$ for MS 0735.6 + 7421, with the remaining sources populating values everywhere between these two extremes. Thus, for cavity radio sources such as those discussed by~\citet{MH07} and \citet{D09a}, there is an enormous range of values of $E_T/(M c^2)$, while for FRIIb sources, this range is much more limited (e.g., \cite{D09a, D22}, see \mbox{Equations (2) and (3))}. 

Surprisingly, even though individual FRI sources input a huge range of values of $E_T/(Mc^2)$ to the ambient 
medium, the total cumulative value of these sources is estimated to be about $3 \times 10^{-3}$ (\cite{DV22, HB23}). This is the nearly identical to that obtained for individual FRIIb sources, which have a relatively  narrow distribution of values for this parameter, as discussed in Section \ref{TotalOutflowEnergy}. Perhaps this can be taken to indicate that most of the energy that heats the circumgalactic medium is input by the most powerful sources, such as \mbox{FRIIb sources. }

Interestingly, the study of FRIIb sources by~\citet{D16} indicates that FRIIb sources are well characterized by Equation (\ref{FL}), with values of 
$A = 0.44 \pm 0.05$ and $B = -1.14 \pm 0.06$, which are consistent with those obtained by~\citet{MH07}. 
Thus, even though the types of radio sources studied are quite different, the general equation that describes 
the kinetic energy carried by the jetted outflow and accretion disk luminosity, each relative to 
the total black hole mass (i.e., written in Eddington units), are quite similar and are described by consistent values of $A$ and $B$. This suggests that a common physical 
mechanism produces jetted outflows from black hole systems~\cite{D18, D19}. However, it is only FRIIb sources that have a well-defined lifetime that can be written as a function of only the luminosity in directed kinetic \mbox{energy ($L$)}. 

Thus, for all of the reasons discussed above, it is unlikely that the well-defined total outflow lifetime 
empirically determined for FRIIb sources is due to a dramatic change in the mass accretion rate that is caused by energy input to the circumgalactic medium by the jetted outflows. 

However, a sequence of processes such as coalescing supermassive black holes that trigger a change in accretion disk properties, which, in turn, either terminates or changes the character of the jetted outflow, 
could explain the properties of FRIIb sources described in Section \ref{TheRadioSources}. In this sense, the accretion disk associated with the primary supermassive black hole could play a role in explaining the remarkable overlap between the properties of FRIIb sources and those required to explain the gravitational wave background.

\section{Summary and Conclusions}
\label{SummaryConclusions}

\subsection{Summary of Key Radio Source Properties}

FRIIb radio sources have unique properties that indicate a particular relationship between the luminosity in 
directed kinetic energy ($L$) and the total outflow lifetime ($T_*$; see Equation (1)). The total outflow lifetime can be combined with the luminosity in directed kinetic energy ($L$) to determine the total outflow energy ($E_T$). The empirically determined values for each of these 
quantities span a large range of values~\cite{ODea2009, D22}. 

The total outflow energy can be combined with an independently determined black hole mass to obtain the ratio of the total outflow energy relative to the supermassive black hole mass ($f_M$; see Equations (2) and (3)). This turns out to be a small number with a small dispersion. Combining Equations (1) and (3) suggests that the total outflow lifetime can be written as $T_7 \simeq 30 ~M_8^{-1}$ (see Equation (5)) and that the luminosity in directed kinetic energy can be written as 
$L_{46} \simeq 0.65 ~M_9^2$ (see Equation (6)), where $M$ is the black hole mass. Similar results are obtained for all source types, as discussed in {Sections} \ref{TotalOutflowEnergy}--\ref{LvsM-fM}. 
In the event that two black holes are involved, the jetted large-scale outflow is taken to be anchored on and originate from the primary black hole, and the black hole mass is that of the primary \mbox{black hole.}

The relationship between $T_*$ and $M$ can also be studied directly, as illustrated by Figure \ref{fig1:Fig1} and discussed in Section \ref{TvsM}. Black hole masses  for some sources have significant uncertainties (see Figure \ref{fig1:Fig1}). The best fit line obtained for the sample of 55 HEG studied and illustrated with Figure \ref{fig1:Fig1} indicates that results obtained by a direct fit to the data for HEG sources are consistent with 
the result of $T_7 \simeq 30 ~M_8^{-1}$ discussed above, and HEG comprise the bulk of the population of studied sources (see the Figure \ref{fig1:Fig1} caption). It is 
possible that there is some variation of $T_*(M)$ by AGN nuclear spectroscopic type, though the uncertainties are large and it is possible that there are no variations by spectroscopic type. 
 \mbox{Equations (5) and (6)} indicate that $T_* \propto M^{-1}$ for all source types, though values of $f_M$ for source types other than HEG have a broader dispersion than HEG sources (see Table 3, column 9 of~\cite{D22}). Surprisingly, the quantity of $f_M$ has no evolution with redshift~\cite{D22}, which also suggests that this is a fundamental quantity that relates the properties of FRIIb radio sources with the properties of the supermassive black hole. 

The radio source properties discussed above indicate a tight connection between the mass of the primary supermassive black hole and the 
properties of the large-scale dual jetted outflow produced by the black hole system. The potential implications for black hole physics explored in this paper are summarized below. 

\subsection{Summary of Connections to Black Hole Physics}

(1) As discussed in Section \ref{Changes-in-BH-Properties}, if the jetted outflow is powered by black hole spin mass-energy extraction, the jetted outflow could terminate when the extracted spin mass energy has the values indicated by Equation (2) or (3). This would indicate that the process shuts down after 
only a small fraction of the spin mass energy has been extracted, since the FRIIb sources studied here have 
a substantial amount of spin mass energy~\cite{D22,D09a,D11}. The remaining spin mass energy could be converted to irreducible black hole mass during the extraction process or the process could shut down after only a small fraction of the available spin mass energy has been tapped. 

In other words, this could indicate that the process shuts down after only a small fraction of the spin mass energy has been extracted while another small fraction has been converted to irreducible black hole mass. This would leave the black hole with a slightly diminished spin mass energy, as in the process discussed by~\cite{R25}, which seems typical for electromagnetic energy extraction mechanisms~\cite{Ketal21, Setal21, Tetal21, RR23,RR24}. This could also occur if most of the spin mass energy is converted to irreducible mass and only a small fraction is extracted to power jets, which would indicate that the extraction process is relatively inefficient, having an efficiency of a few $\times 10^{-3}$, as indicated by Equation (2) or (3). Given the empirically determined value of $f_M$ 
indicated  by Equation (3), mechanisms that would lead to these values could be considered. Obtaining values 
for these quantities requires detailed numerical simulations that depend 
upon the specific spin mass-energy extraction mechanism~\cite{Ketal21, Setal21, Tetal21, RR23, RR24, R25}. 

In this scenario, the jetted outflow is powered by spin mass-energy extraction, and the extraction mechanism must turn off when the value of the extracted spin mass energy reaches the the empirically determined value of $f_M$. This could help to identify the spin mass-energy extraction mechanism. A mechanism that either has an efficiency equal to the empirically determined value of $f_M$ or one that terminates when the empirically determined value of $f_M$ is reached but long before most of the spin mass energy has been extracted or converted to black hole irreducible mass could explain the empirically determined value of $f_M$. 

(2) Perhaps the most interesting possibility is that FRIIb radio sources are produced during the final phases of dual black hole coalescence, during which time the low-frequency gravitational wave background is produced. The jetted outflow would be originating from the primary black hole, and the outflow would terminate when the binary black hole separation shrinks to a particular value, disrupting the jetted outflow so as to modify the radio source type or to terminate the jetted outflow entirely. 
The timescale suggested by 
\cite{KHF25,Nanograv23b} for the 
dual black hole coalescence is remarkably similar to the empirically determined FRII outflow lifetime, as 
discussed in Section \ref{BH Coalescence}. The supermassive black hole properties required to  explain the recently detected nanohertz gravitational wave background, such as the range of black hole masses and the source redshifts, have significant  overlap with the empirically determined properties of FRIIb radio sources. 

In this scenario, FRIIb sources would represent the fraction of coalescing dual black holes that produce powerful, large-scale, jetted outflows, which likely involve dynamically important magnetic fields and a particular range of mass ratios 
between the coalescing black holes. In this case, the number of similar sources at a given redshift contributing to the gravitational wave background could be empirically estimated using the ratio of the age of the universe to the total outflow lifetime as a function of redshift, which is illustrated by Figure \ref{fig1:Fig2} and discussed in Section~\ref{TuoTo}. This would obviate the need to assume a detailed model for the density of coalescing supermassive black holes of a similar type as a function of redshift. 

(3) A third possibility is that the energy input to the circumgalactic medium filters down to cause the mass accretion rate to decrease with a timescale given by Equation (1).  For a variety of reasons discussed in Section~\ref{Changes-in-AD-Properties}, this seems to be the least likely cause for the tight connection between jetted outflow properties and black hole properties for \mbox{FRIIb sources.} 

A change in accretion disk properties 
could, however, indirectly cause the termination of the jetted outflow, or removal from the FRIIb category. If the accretion disk properties of the primary supermassive black hole 
are significantly modified by the presence of a secondary supermassive black hole that is in the process of merging with the primary supermassive black hole, this could explain both the shutting off (or modification) 
of the jetted outflow from the primary black hole and the coincidence of the FRIIb lifetime and the coalescence timescale.

\vspace{6pt}

\funding{This
 research received no external funding.}

\dataavailability{The original data presented in this study are openly available in \cite{Daly2010, ODea2009, WDW97a, WDW97b, 
DDetal08,Detal09, GDW2000, DG02,D22, D19,D09a, DS14, DDO24, D21,D18,D11}.}

\acknowledgments{It is a pleasure to 
thank Ruth Durrer for numerous helpful discussions related to this work and 
my collaborators in these endeavors, especially Stefi Baum, George Djorgovski, Megan Donahue, Eddie Guerra, Daryl Haggard, Preeti Khard, Anan Lu, Matt Mory, Chris O'Dea, Biny Sebastian, Lin Wan, and Greg Wellman. I would also like to thank the anonymous referees for helpful comments and suggestions.
}

\conflictsofinterest{The author declares no conflicts of interest.} 

\begin{adjustwidth}{-\extralength}{0cm}

\reftitle{References}


\PublishersNote{}
\end{adjustwidth}
\end{document}